\documentclass[conference,letterpaper]{IEEEtran}

\usepackage{geometry}
\usepackage[utf8]{inputenc} 
\usepackage[T1]{fontenc}
\usepackage{url}
\usepackage{ifthen}
\usepackage{cite}
\usepackage[cmex10]{amsmath} % Use the [cmex10] option to ensure complicance
\usepackage{graphicx} % Required for inserting images
\usepackage{amssymb, amsthm, extpfeil}
\usepackage{amsfonts,mathtools,mathrsfs,mathtools,color}
\usepackage{hyperref}

\renewcommand{\theequationdis}{{\normalfont (\theequation)}}
\renewcommand{\theIEEEsubequationdis}{{\normalfont (\theIEEEsubequation)}}

\newcommand{\<}{\negmedspace{}} % negative version of \>
\newcommand{\mmse}{\textnormal{\textsf{mmse}}}
\newcommand{\snr}{\textnormal{\textsf{snr}}}

\newtheorem{theorem}{Theorem}
\newtheorem{lemma}{Lemma}
\newtheorem{corollary}{Corollary}

\begin{document}
\title{Differential Properties of Information in Jump-diffusion Channels} 

\author{
  \IEEEauthorblockN{Luyao Fan, Jiayang Zou, Jiayang Gao, and Jia Wang}
  \IEEEauthorblockA{Department of Electronic Engineering\\ 
                    Shanghai Jiao Tong University\\
                    Shanghai, China\\
                    Email: fanluyao@sjtu.edu.cn, qiudao@sjtu.edu.cn, gjy0515@sjtu.edu.cn, jiawang@sjtu.edu.cn}
}

\maketitle

\begin{abstract}    
    We propose a channel modeling using jump-diffusion processes, and study the differential properties of entropy and mutual information.
    By utilizing the Kramers-Moyal and Kolmogorov equations, we express the mutual information between the input and the output in series and integral forms, presented by Fisher-type information and mismatched KL divergence. We extend de Bruijn's identity and the I-MMSE relation to encompass general Markov processes.
    \vspace{-3pt}
\end{abstract}
\section{Introduction}
\vspace{-3pt}
Measures in information theory, such as Shannon entropy and mutual information~\cite{thomas2006elements}, are closely related to statistical measures like Fisher information and other Fisher-type information~\cite{bobkov2024fisher}, as well as to estimators such as maximum likelihood estimation and Bayesian estimation. Researchers have devoted considerable efforts to uncovering the various interconnections between them. In discrete cases, Fano's inequality~\cite{thomas2006elements} provides an upper bound on the error rate determined by entropy. However, in continuous cases, finding such connections is more challenging. De Bruijn's identity~\cite{Dbj} states that the time derivative of the entropy of a channel output with Gaussian noise is equal to its Fisher information (cf. equation (\ref{Dbj})). Following this, McKean et al.~\cite{mckean1966speed} demonstrated that the aforementioned entropy is concave and hypothesized that the sign of the derivatives of entropy alternates with each order. This hypothesis was subsequently confirmed up to the fourth order by Cheng et al.~\cite{cheng2015higher}.

The proposal of the I-MMSE identity by Guo et al.~\cite{agn} established a connection between entropy and statistical estimation quantities. This identity states that the derivative of mutual information between the channel input and output with respect to the signal-to-noise ratio (SNR)~\cite{thomas2006elements} is proportional to the minimum mean square error (MMSE) in estimating the input from the output (cf. Lemma \ref{thm:immse}). Subsequently, this identity was extended to other channels, such as Poisson channels\cite{atar2012mutual, jiao2013pointwise} and discrete-time Lévy channels~\cite{jiao2014relations}. Wibisono et al.~\cite{fpc} further extended the identity to Fokker-Planck channels, which are characterized by stochastic differential equations (SDEs). They revealed the relationships between mutual information and mutual Fisher-type information (weighted or generalized mutual Fisher information) (cf. Lemma \ref{thm:diff_I_dp}), and the identity between Fisher-type information and weighted MMSE (generalized MMSE) (cf. Lemma \ref{thm:j_mmse}).

With the widespread adoption of mobile communications, channel characteristics are increasingly shaped by factors such as location, time, and nonlinear effects. This growing complexity necessitates more precise modeling approaches. To address this, we utilize continuous-time Markov processes with continuous state spaces, offering enhanced accuracy and sophistication. In particular, the jump-diffusion process is employed to model a broad range of Markov channels, excluding certain special cases~\cite{hanson2007applied}. Additionally, from a practical application viewpoint, with drift describing deterministic evolution, diffusion introducing continuous small-scale noise, and jumps modeling abrupt large-scale distortions, the jump-diffusion channel represents a more refined model that meets modern needs. Similar modeling is widely applied in fields including physics \cite{honisch2012extended, daly2006probabilistic,lehnertz2018characterizing}, biology \cite{anvari2016disentangling, sirovich2013cooperative}, and finance \cite{tankov2003financial, da2019jump}.

Building on the work of Wibisono et al. \cite{fpc}, we extend the differential properties of information to jump-diffusion channels. Utilizing the Kramers-Moyal equations, we provide a series expansion of entropy and mutual information in terms of Fisher-type information. The Kramers-Moyal coefficients (propagator moment functions) in this expansion are more straightforward to estimate from time series data with a finite sampling intervals~\cite{rydin2021arbitrary}, making it convenient for computing approximate solutions. Furthermore, by employing the Kolmogorov equations~\cite{hanson2007applied}, we demonstrate that the time derivative of mutual information in the jump-diffusion channel equals the sum of the respective derivatives in the corresponding diffusion and jump channels. Here, the diffusion part corresponds to Fisher-type information, while the jump part can be represented as the expectation of mismatched KL divergence~\cite{verdu2010mismatched}, both of which are non-negative. We also focus our results on additive noise channels, which exhibit some unique properties. The I-MMSE identities in Fokker-Planck channels and additive Gaussian channels can be recovered as special cases of our results.

The remainder of the paper is organized as follows: Section \ref{bkg} reviews existing results and analyzes the properties of jump-diffusion processes. Section \ref{rst} presents the series and integral forms of the time derivative and discusses the special properties of additive jump-diffusion channels. We conclude with Section \ref{dscs}, which includes a discussion of open problems. Appendices \ref{proof_lem}, \ref{proof_thm}, and \ref{proof_coro} provide detailed proofs of lemmas, theorems, and corollaries.

\section{Preliminaries of jump-diffusion channels}\label{bkg}
\subsection{Jump-diffusion channels}
Consider a jump-diffusion channel that outputs a real-valued stochastic process \((X_t)_{t\geqslant 0}\) following the SDE
\begin{IEEEeqnarray}{c}\label{SDE_jdp}
    dX_t=a( X_t,t ) dt+\sqrt{b( X_t,t )}dW_t+\varXi ( X_t,t ) dN_t
\end{IEEEeqnarray}
where \((W_t)_{t\geqslant 0}\) is standard Brownian motion, \(\varXi \sim w( \xi | X_t,t ) \) is a real-valued random variable representing the jump size, and \((N_t)_{t\geqslant 0}\) is Poisson process~\cite{borodin2017stochastic} with intensity \(\lambda ( X_t,t )\).
\(a( x,t )\), \(b( x,t )\geqslant 0\), \(\lambda ( x,t )\geqslant 0 \) and \(w( \xi |x,t )\geqslant 0\) are assumed to be smooth functions with respect to all variables for simplicity of analysis.
The drift \(a( x,t ) dt\) is deterministic, the diffusion \(\sqrt{b( x,t )}dW_t\) introduces continuous small-scale noise, and the jump \(\varXi ( X_t,t ) dN_t\) models abrupt large-scale distortion.

For small \(\Delta t>0\), we define the propagator \(\Delta X_t\) by
\begin{IEEEeqnarray}{rCl}\label{SDE_deltat}
    \!\!\!\Delta X_t&\!=\!&X_{t+\Delta t}- X_t\nonumber
    \\
    & \!=\! & a( X_t,\!t )\Delta t\!+\!\sqrt{b( X_t,\!t )\Delta t}Z\!+\!\varXi ( X_t,\!t ) Y\!\! +\! \tilde{\mathcal{O}}(\Delta t^2)
\end{IEEEeqnarray}
where \(Z\sim \mathcal{N}(0,1)\) is the standard normal distribution, and \(Y\sim\mathrm{B}(1,\lambda ( X_t,t )\Delta t)\) is the Bernoulli distribution taking value 1 with probability \(\lambda ( X_t,t )\Delta t\), and \(\tilde{\mathcal{O}}(\Delta t^2)\) denotes that if \(U=\tilde{\mathcal{O}}(\Delta t^2)\), \(1-\mathbb{P}(U\ne\mathcal{O}(\Delta t^2))=\mathcal{O}(\Delta t^2)\).

The jump-diffusion process is versatile enough to effectively model a wide range of Markov processes, with the exception of certain Lévy-driven processes characterized by infinite jump rates and nonlinear, time-dependent coefficients~\cite{hanson2007applied}.

\begin{lemma}\label{lem:jdc_PDF}
    The probability density function (PDF) of the propagator \(\Delta X_t\) in (\ref{SDE_deltat}) is
    \begin{align*}
        p_{\Delta X} ( \xi |\Delta t;x,t ) = w&( \xi |x,t ) \lambda ( x,t ) \Delta t
        \\
        +\frac{1-\lambda ( x,t ) \Delta t}{\sqrt{2\pi b( x,t ) \Delta t}}&\exp \left( -\frac{( \xi -a( x,t ) \Delta t ) ^2}{2b( x,t ) \Delta t} \right)
        +\mathcal{O}(\Delta t^2).
    \end{align*}
    \begin{proof}
        Briefly, the cross terms arising from the jump and diffusion components are higher-order infinitesimals of \(\Delta t\). For a detailed proof, please refer to Appendix \ref{app:lem:jdc_PDF}. 
    \end{proof}
\end{lemma}

\subsection{Equations of PDFs}
\subsubsection{Fokker-Planck equation}
When there is only diffusion, i.e. \(\varXi ( x,t )\lambda(x,t)\equiv 0\), the SDE can be written as
\begin{IEEEeqnarray}{c}\label{SDE_dp}
    dX_t=a( X_t,t ) dt+\sqrt{b( X_t,t )}dW_t.
\end{IEEEeqnarray}
\begin{lemma}
    The PDF \(p(x,t)\) of the diffusion in (\ref{SDE_dp}) satisfies the Fokker-Planck equation \cite{mkp}
    \begin{equation}\label{fpe}
        \frac{\partial p( x,t )}{\partial t}=-\frac{\partial}{\partial x}( a( x,t ) p( x,t ) ) +\frac{1}{2}\frac{\partial ^2}{\partial x^2}( b( x,t ) p( x,t ) ).
    \end{equation}
\end{lemma}

\subsubsection{Master equation}
When there is only jump, i.e. \(a(x,t)\equiv b(x,t)\equiv 0\), the SDE can be written as
\begin{IEEEeqnarray}{c}\label{SDE_jp}
    dX_t= \varXi ( X_t,t ) dN_t.
\end{IEEEeqnarray}
\begin{lemma}
    The PDF \(p(x,t)\) of the jump process in (\ref{SDE_jp}) satisfies the master equation \cite{mkp}
    \begin{IEEEeqnarray}{rCl}\label{me}
        \frac{\partial p( x,t )}{\partial t}&=&\int_{\mathbb{R} }{p( x-\xi ,t ) \lambda ( x-\xi ,t ) w( \xi |x-\xi ,t ) d\xi}\nonumber
        \\
        &&\<-p( x,t ) \lambda ( x,t ).
    \end{IEEEeqnarray}
\end{lemma}

\subsubsection{Kramers-Moyal equation}
\begin{lemma}\label{lem:kme}
    The PDF \(p(x,t)\) of \(X_t\) in (\ref{SDE_jdp}) satisfies the Kramers-Moyal equation~\cite{mkp}
    \begin{IEEEeqnarray*}{c}
        \frac{\partial p( x,t )}{\partial t}=\sum_{n=1}^{\infty}{\frac{( -1 ) ^n}{n!}\frac{\partial ^n}{\partial x^n}\left[ B_n( x,t ) p( x,t ) \right]}
    \end{IEEEeqnarray*}
    where
    \begin{IEEEeqnarray}{c}\label{pmf}
        B_n( x,t ) =\lim_{\Delta t\rightarrow 0} \frac{\mathbb{E} \left[ \Delta X _t^n|X_t=x \right]}{\Delta t}
    \end{IEEEeqnarray}
    is the \(n^{\text{th}}\) propagator moment function (Kramers-Moyal coefficient) of the stochastic process \((X_t)_{t\geqslant 0}\).
    \begin{proof}
        While this lemma is proven in~\cite{mkp}, we offer a simple alternative proof in Appendix \ref{app:lem:kme}.
    \end{proof}
\end{lemma}
The value of \(B_n(x,t)\) can be estimated from time series data with a finite sampling intervals~\cite{rydin2021arbitrary}, and can also be derived from the functions in the SDE (\ref{SDE_jdp}).
\begin{theorem}\label{thm:kmc}
    \begin{IEEEeqnarray*}{c}
        B_n( x,t )=\begin{cases}
            \lambda ( x,t ) w_1( x,t ) +a( x,t ) ,&n=1 \\
            \lambda ( x,t ) w_2( x,t ) +b( x,t ) ,&n=2 \\
            \lambda ( x,t ) w_n( x,t ) ,&n\geqslant 3  \\
        \end{cases}
    \end{IEEEeqnarray*}
    where
    \begin{IEEEeqnarray*}{c}
        w_n( x,t ) =\int_{\mathbb{R}}{\xi ^nw( \xi |x,t ) d\xi}
    \end{IEEEeqnarray*}
    is the \(n^\text{th}\) moment of \(w(x,t)\).
    \begin{proof}
        In brief, while the higher-order moments of the diffusion component are higher-order infinitesimals of \(\Delta t\), those of the jump component persist. For a detailed proof, please refer to Appendix \ref{app:thm:kmc}. 
    \end{proof}
\end{theorem}

\subsubsection{Kolmogorov equation}
Not all sets of propagator moment functions \(\{B_n(x,t)\}\) are possible to generate a Markov process~\cite{PhysRev.162.186, problemOfMoments}.
Thus to ensure the equation describing the process \((X_t)_{t\geqslant 0}\) to be possible, we can rewrite the time derivative of the probability in the form of the functions in the SDE (\ref{SDE_jdp}) instead of propagator moment functions.
\begin{theorem}\label{thm:fpme}
    The PDF \(p(x,t)\) of \(X_t\) satisfies the Kolmogorov equation~\cite{hanson2007applied}
    \begin{IEEEeqnarray}{r}\label{thm:eq_fpme}
        \frac{\partial p( x,t )}{\partial t} = \int_{\mathbb{R}}{p( x-\xi ,t ) \lambda ( x-\xi ,t ) w( \xi |x-\xi ,t ) d\xi}\nonumber
        \\
        -p( x,t ) \lambda ( x,t )\hspace{4.2cm}\nonumber
        \\
        -\frac{\partial}{\partial x}( a( x,t ) p( x,t ) ) +\frac{1}{2}\frac{\partial ^2}{\partial x^2}( b( x,t ) p( x,t ) ).
    \end{IEEEeqnarray}
 
    \begin{proof}
        The SDE we have defined differs slightly from the one presented in~\cite{hanson2007applied}, so a specific proof is provided in Appendix \ref{app:thm:fpme}.
    \end{proof}
\end{theorem}
The first two items of the RHS in (\ref{thm:eq_fpme})
\begin{IEEEeqnarray*}{c}
    \int_{\mathbb{R}}{p( x - \xi ,t ) \lambda ( x-\xi ,t ) w( \xi |x-\xi ,t ) d\xi}-p( x,t ) \lambda ( x,t )
\end{IEEEeqnarray*}
is the RHS of the master equation (\ref{me}) of a jump Markov process (\ref{SDE_jp}), and the rest part
\begin{IEEEeqnarray*}{c}
    -\frac{\partial}{\partial x}( a( x,t ) p( x,t ) ) +\frac{1}{2}\frac{\partial ^2}{\partial x^2}( b( x,t ) p( x,t ) )
\end{IEEEeqnarray*}
is the RHS of the Fokker-Planck equation (\ref{fpe}) of a diffusion (\ref{SDE_dp}), which indicates that the jump part and diffusion part of a jump-diffusion process can be studied separately in a small time interval. Indeed, Theorem \ref{thm:diff_h} and Theorem \ref{thm:diff_i} confirm this indication.

\subsection{De Bruijn’s identity and I-MMSE}
Consider the additive Gaussian noise channel
\begin{IEEEeqnarray}{c}\label{Eq:Gaussian}
    X_t=X_0+\sqrt{t}Z
\end{IEEEeqnarray}
where \(Z\sim\mathcal{N}(0,1)\) is independent of \(X_0\). De Bruijn’s identity \cite{Dbj} asserts that
\begin{IEEEeqnarray}{c}\label{Dbj}
    \frac{d}{dt}h(X_t)=\frac{1}{2}J(X_t)
\end{IEEEeqnarray}
where \(h(X_t)=-\mathbb{E}[\log p_{X_t}(X_t)]\) is the Shannon entropy with
\(p_{X_t}\) denoting the PDF of \(X_t\), and
\begin{IEEEeqnarray*}{c}
    J(X_t) = \mathbb{E} \biggl[ \Bigl(\frac{\partial}{\partial x}\log p_{X_t}(X_t)\Bigr)^2 \biggr] =  \int_{\mathbb R} \frac{p_{X_t}'(x)^2}{p_{X_t}(x)} dx > 0
\end{IEEEeqnarray*}
is the (nonparametric) Fisher information.

Another common parameterization of the channel (\ref{Eq:Gaussian}) is
\begin{IEEEeqnarray*}{c}
    Y_ \snr = \sqrt{\snr} X + Z
\end{IEEEeqnarray*}
where $\snr > 0$ is the signal to noise ratio and $Z \sim \mathcal{N}(0,1)$ is independent of $X$.
Guo et al. \cite{agn} established the following I-MMSE relation, which states that the mutual information \(I(\snr) = I(X; Y_\snr)\) increases at a rate given by the MMSE, and showed their result is equivalent to de Bruijn's identity \cite{Dbj}.

\begin{lemma}[Guo et al. \cite{agn}]\label{thm:immse}
    \begin{IEEEeqnarray*}{c}
        \frac{d}{d\snr} I(\snr) = \frac{1}{2} \mmse (X \,|\, Y_\snr),
    \end{IEEEeqnarray*}
    where \(\mmse(X\,|\,Y_\snr) = \mathbb{E}[(X - \mathbb{E}[X \,|\, Y_\snr])^2]\) denotes the minimum mean square error for estimating \(X\) from \(Y_\snr\).
\end{lemma}
Wibisono et al. \cite{fpc} showed the aforementioned in terms of the time parameterization (\ref{Eq:Gaussian}). By setting $\snr = \frac{1}{t}$ and $X_t = \sqrt{t} Y_{1/t}$ we see that Theorem \ref{thm:immse} is equivalent to
\begin{IEEEeqnarray*}{c}\label{eq: immse normal}
    \frac{d}{dt} I(X_0;X_t) = -\frac{1}{2t^2} \mmse (X_0 \,|\, X_t).
\end{IEEEeqnarray*}
Wibisono et al. \cite{fpc} extended the I-MMSE relation to Fokker-Planck channels that outputs a diffusion process \((X_t)_{t\geqslant 0}\) following the SDE (\ref{SDE_dp}).
\begin{lemma}[Wibisono et al. \cite{fpc}]\label{thm:diff_h_dp}
    \begin{IEEEeqnarray*}{c}
        \frac{d}{dt} h(X_t) = \frac{1}{2} J_{b_t}(X_t) + \mathbb{E}\biggl[\frac{\partial}{\partial x}a(X_t,t) - \frac{1}{2} \frac{\partial^2}{\partial x^2} b(X_t,t)\biggr]
    \end{IEEEeqnarray*}
    where the Fisher-type information with respect to \(b:\mathbb{R}\to\left[0,+\infty\right)\) is
    \begin{IEEEeqnarray}{rCl}\label{fishertype}
        J_b(X)&=&\mathbb{E}\biggl[b(X)\Bigl(\frac{\partial}{\partial x}\log p_X(X)\Bigr)^2\biggr]\nonumber
        \\
        &=&\int_\mathbb{R}{b(x)\frac{p_X^2(x)}{p_X(x)}dx}.
    \end{IEEEeqnarray}
\end{lemma}

\begin{lemma}[Wibisono et al. \cite{fpc}]\label{thm:diff_I_dp}
    \begin{IEEEeqnarray*}{c}
        \frac{d}{dt} I(X_0; X_t) = -\frac{1}{2} J_{b_t}(X_0;X_t)
    \end{IEEEeqnarray*}
    where the mutual Fisher-type information with respect to \(b\) is
    \begin{IEEEeqnarray}{rCl}\label{mufishertype}
        J_b(X;Y)&=&J_b(Y|X)-J_b(Y)\nonumber
        \\
        &=&\int_\mathbb{R}{p_X(x)J_b(Y|X=x)dx}-J_b(Y).
    \end{IEEEeqnarray}
\end{lemma}

\begin{lemma}[Wibisono et al. \cite{fpc}]\label{thm:j_mmse}
    \begin{IEEEeqnarray*}{c}
        J_b(X;Y)=\mmse_b(\varphi(X,Y) \,|\, Y)
    \end{IEEEeqnarray*}
    where
    \begin{IEEEeqnarray}{c}\label{gen_mmse}
        \mmse_b(Y|X)= \min_{T} \mathbb{E}[b(X) (T(X) - Y)^2]
    \end{IEEEeqnarray}
    is the MMSE of \(Y\) given \(X\) with respect to \(b\), and
    \begin{IEEEeqnarray}{c}\label{scoreFun}
        \varphi(x,y):=\frac{\partial}{\partial y}\log p_{Y|X}(y|x)
    \end{IEEEeqnarray}
    is the pointwise score function.
\end{lemma}

\section{Main results}\label{rst}
We now generalize the information-estimation relations from the additive Gaussian channel (\ref{Eq:Gaussian}) and the diffusion channel (\ref{SDE_dp}) to the jump-diffusion channel (\ref{SDE_jdp}). We express the time derivatives of entropy and mutual information in terms of series expansion, Fisher-type information and KL divergence, which can be considered as extensions of de Bruijn’s identity (\ref{Dbj}) and I-MMSE relation. The entropy may not increase over time, but the mutual information always decrease at a rate higher than half of the generalized MMSE (\ref{gen_mmse}).
\subsection{Series form of time derivative}
We calculate the series expansion of time derivative of entropy and mutual information in terms of propagator moment functions (\ref{pmf}).
\begin{theorem}\label{thm:KM_diff_h}
    The output entropy of the channel (\ref{SDE_jdp}) satisfies
    \begin{IEEEeqnarray*}{rCl}
        \frac{d}{dt}h( X_t ) &=&\frac{1}{2}J_{B_{2,t}}( X_t )+\mathbb{E} \biggl[ \frac{\partial}{\partial x}B_1( X_t,t ) -\frac{\partial ^2}{\partial x^2}B_2( X_t,t ) \biggr] \nonumber
        \\
        &&\<-\sum_{n=3}^{\infty}{\frac{1}{n!}\mathbb{E} \biggl[ B_n( X_t,t ) \frac{\partial ^n}{\partial x^n}\log p_{X_t}( X_t,t ) \biggr]}
    \end{IEEEeqnarray*}
    where \(J_{B_{2,t}}(X_t)\) refers to the Fisher-type information (\ref{fishertype}) with respect to \(B_2(x,t)\).
    \begin{proof}
        The result is derived through the method of integration by parts, followed by the application of the Kramers-Moyal equation. For a detailed proof, please refer to Appendix \ref{app:thm:KM_diff_h}. 
    \end{proof}
\end{theorem}
Note that Theorem \ref{thm:KM_diff_h} reduces to Lemma \ref{thm:diff_h_dp} when \(B_n\equiv 0\) for \(n\geqslant 3\). More specifically, it is equivalent to de Bruijn’s identity (\ref{Dbj}) if \(B_1\equiv 0\), \(B_2\equiv 1\) and \(B_n\equiv 0\) for \(n\geqslant 3\).
\begin{theorem}\label{thm:KM_diff_i}
    The mutual information between the input \(X_0\) and the output \(X_t\) of the channel (\ref{SDE_jdp}) satisfies
    \begin{align}\label{series}
        \frac{d}{dt}I( X_0;X_t )&=-\frac{1}{2}J_{B_{2,t}}( X_0;X_t ) \nonumber
        \\
        +\sum_{n=3}^{\infty}&{\frac{1}{n!}\mathbb{E} \biggl[ B_n( X_t,t ) \frac{\partial ^n}{\partial x_{t}^{n}}\log p_{X_0|X_t}( X_0|X_t ) \biggr]}
    \end{align}
    where \(J_{B_{2,t}}(X_0,X_t)\) refers to the mutual Fisher-type information (\ref{mufishertype}) with respect to \(B_2(x,t)\).
    \begin{proof}
        Starting with the identity \(\frac{d}{dt}I( X_0;X_t ) = \frac{d}{dt}h( X_t ) -\frac{d}{dt}h( X_t|X_0 )\), we then apply Theorem \ref{thm:KM_diff_h} to obtain this conclusion. For a detailed proof, please refer to Appendix \ref{app:thm:KM_diff_i}. 
    \end{proof}
\end{theorem}
The series terms in (\ref{series}) generally decrease rapidly as \(n\) gets larger, so Theorem \ref{thm:KM_diff_i} introduces an efficient method to estimate time derivative using series expansion. However, the restrictions of \(B_n\) (Theorem \ref{thm:kmc}, \cite{PhysRev.162.186, problemOfMoments}) are not shown explicitly in the expression, so we cannot tell exactly how large the time derivative is, nor even whether it is positive or negative. Thus we will introduce the time derivative in terms of the functions in the SDE (\ref{SDE_jdp}), which may help us determine the value more precisely.

\begin{corollary}\label{KMmmse}
    According to Theorem \ref{thm:KM_diff_i} and Lemma \ref{thm:j_mmse},
\begin{IEEEeqnarray*}{rCl}
    \frac{d}{dt}I( X_0;X_t )&=&-\frac{1}{2}\mmse_{B_{2,t}}(\varphi(X_0,X_t)|X_t) 
    \\
    &+&\<\sum_{n=3}^{\infty}{\frac{1}{n!}\mathbb{E} \left[ B_n( X_t,t ) \frac{\partial ^n}{\partial x_{t}^{n}}\log p_{X_0|X_t}( X_0|X_t ) \right]}
\end{IEEEeqnarray*}
where \(\varphi(\cdot,\cdot)\) is the score function (\ref{scoreFun}).
\end{corollary}

Corollary \ref{KMmmse} reveals the relationship between the time derivative of mutual information and MMSE. \(B_n\) for \(n\geqslant 3\) are generally considered tiny enough in contrast to \(B_2\), so the time derivative is approximately MMSE.

\subsection{Integral form of time derivative}
\begin{theorem}\label{thm:diff_h}
    The output entropy of the channel (\ref{SDE_jdp}) satisfies
    \begin{IEEEeqnarray*}{rCl}
        \frac{d}{dt}h( X_t ) &=& \frac{1}{2}J_{b_t}( X_t ) + \mathbb{E} \biggl[ \lambda ( X_t,t ) \log \frac{p_{X_t}( X_t )}{p_{X_t}( X_t+\varXi )} \biggr]
        \\
        &&\<+\mathbb{E} \biggl[\frac{\partial}{\partial x}a( X_t,t ) -\frac{1}{2}\frac{\partial ^2}{\partial x^2}b( X_t,t ) \biggr] .
    \end{IEEEeqnarray*}
    \begin{proof}
        Analogous to Theorem \ref{thm:KM_diff_h}, the result is derived using the method of integration by parts, followed by the application of the Kolmogorov equation. For a detailed proof, please refer to Appendix \ref{app:thm:diff_h}. 
    \end{proof}
\end{theorem}
Unlike that in Gaussian channels, output entropy in SDE channels may not increase over time due to the arbitrariness of functions in (\ref{SDE_jdp}).
\begin{theorem}\label{thm:diff_i}
    The mutual information between the input \(X_0\) and the output \(X_t\) of the channel (\ref{SDE_jdp}) satisfies
    \begin{IEEEeqnarray}{rCl}\label{integral}
        \frac{d}{dt}I( X_0;X_t ) &=&-\mathbb{E} \left[ \lambda ( X_t,t ) D( p_{X_0|X_t}\parallel p_{X_0|X_{t,\varXi}}|\varXi,X_t ) \right] \nonumber
        \\
        &&\<-\frac{1}{2}J_{b_t}( X_0;X_t )
    \end{IEEEeqnarray}
    where \(p_{X_0|X_{t,\xi}}\!:=p_{X_0|X_t}(x_0|x_t+\xi)\) refers to the spatial shift \(\xi\) of the PDF \(p_{X_0|X_t}(x_0|x_t)\) with respect to \(x_t\), and 
    \begin{align*}
        D( p_{X_0|X_t}\!\parallel\;& p_{X_0|X_t,\varXi}|\varXi =\xi,X_t=x_t )
    \\
        &=\int_{\mathbb{R}}{p_{X_0|X_t}( x_0|x_t ) \log \frac{p_{X_0|X_t}( x_0|x_t )}{p_{X_0|X_t}( x_0|x_t+\xi )}dx_0}
    \end{align*}
    is the KL divergence between \(p_{X_0|X_t}\) and \(p_{X_0|X_{t,\varXi}}\) given \(\varXi=\xi\) and \(X_t=x_t\).
    \begin{proof}
        Starting with the identity \(\frac{d}{dt}I( X_0;X_t ) = \frac{d}{dt}h( X_t ) -\frac{d}{dt}h( X_t|X_0 )\), we apply Theorem \ref{thm:diff_h}, and finally rely on the proof of the conclusion "Conditioning increases divergence" in \cite{Polyanskiy_Wu_2025} to complete the proof. For a detailed proof, please refer to Appendix \ref{app:thm:diff_i}. 
    \end{proof}
\end{theorem}
The first part of the RHS in (\ref{integral}) represents the expectation of mismatched KL divergence, which is attributed to the jump process, while the second part represents Fisher-type information, attributed to the diffusion process.

Theorem \ref{thm:KM_diff_i} and Theorem \ref{thm:diff_i} are both expressions of time derivatives. Comparatively, the series form Theorem \ref{thm:KM_diff_i} is more suitable for approximate estimation~\cite{lehnertz2018characterizing}, while the integral form Theorem \ref{thm:diff_i} is better suited for exact calculation.
\begin{corollary}\label{cor:I-mmse}
    \begin{IEEEeqnarray*}{rCl}
        \frac{d}{dt}I( X_0;X_t ) &\leqslant& -\frac{1}{2}J_{b_t}( X_0;X_t ) 
        \\
        &=& -\frac{1}{2}\mmse_{b_t}(\varphi(X_0,X_t)|X_t).
    \end{IEEEeqnarray*}
\end{corollary}
It follows that the mutual information between the input and the output of a jump-diffusion channel is always decreasing over time, which can be regarded as data processing inequality~\cite{thomas2006elements} in continuous-time cases.

\subsection{Additive noise channels}
We specialize our results on additive noise channels here. In such channels, the Kramers-Moyal equation of the output can yield an explicit solution, and entropy of the output is always increasing.
The jump-diffusion process (\ref{SDE_jdp}) is state-homogeneous if the propagator is independent of the current state \(X_t\), corresponding to the additive channel. Then the SDE can be written as
\begin{IEEEeqnarray}{c}\label{SDE_add}
    dX_t=a( t ) dt+\sqrt{b( t )}dW_t+\varXi ( t ) dN_t
\end{IEEEeqnarray}
where
\(
    \lambda ( t ) = \lambda ( x,t ) , w( \xi ,t ) = w( \xi |x,t ) , a( t ) =a( x,t ) ,\\b( t ) =b( x,t )
\)
according to the state-homogeneity.

We can simply apply the integration to get
\begin{IEEEeqnarray}{c}
    X_t=X_0+\int_0^t{a(\tau)d\tau+\sqrt{b(\tau)}dW_\tau+\varXi(\tau) dN_\tau }
\end{IEEEeqnarray}
where the integration
\begin{IEEEeqnarray}{c}
    Z_t=\int_0^t{a(\tau)d\tau+\sqrt{b(\tau)}dW_\tau+\varXi(\tau)dN_\tau}
\end{IEEEeqnarray}
is independent of \(X_0\), so the process is equivalent to an additive noise channel
\begin{IEEEeqnarray}{c}
    X_t = X_0 + Z_t
\end{IEEEeqnarray}
where the noise \((Z_t)_{t\geqslant 0}\) is a state-homogeneous process with the initial value \(Z_0=0\), i.e. \(p_Z(x,0)=\delta(x)\).
\begin{corollary}\label{crl:add_PDF}
    The PDF \(p(x,t)\) of the state-homogeneous jump-diffusion process in (\ref{SDE_add}) satisfies
        \begin{IEEEeqnarray*}{l}
            p(x,t)=\dfrac{1}{2\pi}\int_{\mathbb{R}}
            \left(
            \int_{\mathbb{R}} p(x,0)e^{-ikx} dx
            \right)
            \\\qquad\qquad
            \cdot \exp\left(
            \int_{0}^{t}\left[
            \left(
            \int_{\mathbb{R}} w(x,\tau)e^{-ikx}dx -1
            \right)\lambda(\tau)\right.\right. 
            \qquad\\
            \IEEEeqnarraymulticol{1}{r}{\left.\left.-ika(\tau)-\frac{k^2}{2}b(\tau)
            \right]d\tau +ikx
            \right)dk.}
        \end{IEEEeqnarray*}

    \begin{proof}
        The result is obtained by appling the Fourier transform to the Kolmogorov equation. For a detailed proof, please refer to Appendix \ref{app:crl:add_PDF}. 
    \end{proof}
\end{corollary}
Corollary \ref{crl:add_PDF} is applicable to the PDF of both \(X_t\) and \(Z_t\): \(p(x,t)=p_{X_t}(x)\) if \(p(x,0)=p_{X_0}(x)\), and \(p(x,t)=p_{Z_t}(x)\) if \(p(x,0)=\delta(x)\).
\begin{corollary}\label{crl:diff_h_add}
    For additive noise channel (\ref{SDE_add}), the output entropy satisfies
    \begin{IEEEeqnarray*}{c}
        \frac{d}{dt}h( X_t ) = \lambda ( t ) \mathbb{E} \left[ D( p_{X_t}\parallel p_{X_{t,\varXi}} ) \right] +\frac{1}{2}J_{b_t}( X_t )
    \end{IEEEeqnarray*}
    where \(p_{X_{t,\xi}}\!:=p_{X_t}(x+\xi)\) refers to the spatial shift \(\xi\) of the PDF \(p_{X_t}(x)\).
    \begin{proof}
        This corollary is a simplified version of Theorem \ref{thm:diff_h}, as the derivatives of \(a\) and \(b\) with respect to \(x\) are zero in the additive case. For a detailed proof, please refer to Appendix \ref{app:crl:diff_h_add}.
    \end{proof}
\end{corollary}
Because of state homogeneity, output entropy in additive noise channels is always increasing, which is consistent with intuition.

Similarly, we have specialized results for mutual information in additive noise channels:
\begin{corollary}
For additive noise channel (\ref{SDE_add}), the mutual information between the input and the output satisfies
        \begin{IEEEeqnarray*}{rCl}
        \frac{d}{dt}I( X_0;X_t ) &=&-\lambda ( t )D( p_{X_0|X_t}\parallel p_{X_0|X_{t,\varXi}}|p_{X_t} )  
        \\
        &&\<-\frac{1}{2}J_{b_t}( X_0;X_t )
    \end{IEEEeqnarray*}
    where \(
    D(p_{Y|X}\!\parallel\! q_{Y|X}|p_X):=\mathbb{E}_{x\sim p_X}[D(p_{Y|X=x}\!\parallel\! q_{Y|X=x})]
    \) is the conditional KL divergence~\cite{Polyanskiy_Wu_2025}.
\end{corollary}
    
\section{Discussion and future work}\label{dscs}
We have extended the I-MMSE identities to general Markov processes, encompassing both diffusion and jumps. According to the data processing inequality~\cite{thomas2006elements}, mutual information in such Markov processes always decreases over time. We provide the exact rate of this decrease in Theorem \ref{thm:KM_diff_i} and Theorem \ref{thm:diff_i}.

The series expansion in Theorem \ref{thm:KM_diff_i} includes different orders of Fisher-type information, which have not been discussed in \cite{bobkov2024fisher}. We are interested in exploring the characteristics of these quantities.

The mismatched KL divergence in Theorem \ref{thm:diff_i} is similar to the concepts discussed in \cite{verdu2010mismatched, jiao2013pointwise}, but not exactly the same. We are intrigued by the deeper meaning behind this expression.

Corollary \ref{cor:I-mmse} demonstrates that mutual information decreases at a rate higher than that of the generalized MMSE (\ref{gen_mmse}). However, the MMSE may not be suitable for describing the mismatched KL divergence term, as it is not Gaussian. Jiao et al.~\cite{jiao2013pointwise} found a loss function suitable for Poisson channels, leading us to conjecture that a similar loss function may be applicable to jump-diffusion channels. This conjecture, if proven, would transform the current inequality into an equality between information and estimation.

The Hamburger moment problem \cite{problemOfMoments} raises questions about the criterion that a distribution can be determined by different orders of moments. We are intrigued by a similar issue: whether a set of propagator moment functions in the Kramers-Moyal equation can generate a Markov process. Pawula \cite{PhysRev.162.186} discovered that the propagator moment functions either truncate at the second term or remain positive for all even terms. However, this constraint is not sufficient. If the sufficient and necessary criteria for generating a Markov process can be explicitly defined, we might be able to derive new insights into the variation rate of mutual information based on Theorem \ref{thm:KM_diff_i}.
\bibliographystyle{IEEEtran}
\bibliography{bibliofile}

\newpage
\onecolumn
\appendices

\section{Proofs of lemmas}\label{proof_lem}
\subsection{Proof of Lemma \ref{lem:jdc_PDF}}\label{app:lem:jdc_PDF}
According to the definition of Brownian motion, the PDF of the diffusion propagator \(a( X_t,t )\Delta t+\sqrt{b( X_t,t )\Delta t}Z\) is
\begin{IEEEeqnarray*}{c}
    p _d( \xi |\Delta t;x,t ) = \frac{1}{\sqrt{2\pi b( x,t ) \Delta t}}\exp \left( -\frac{\left( \xi -a( x,t ) \Delta t \right) ^2}{2b( x,t ) \Delta t} \right)+\mathcal{O}(\Delta t^2)
\end{IEEEeqnarray*}
and according to the definition of Poisson process\cite{borodin2017stochastic}, the probability of jumping to another state is approximately \(\lambda ( x,t ) \Delta t\), and that of staying in the current state is \(( 1-\lambda ( x,t ) \Delta t )\), so the PDF of the jump propagator \(\varXi ( X_t,t ) Y\) is
\begin{IEEEeqnarray*}{c}
    p_j( \xi |\Delta t;x,t ) = w( \xi |x,t ) \lambda ( x,t ) \Delta t+( 1-\lambda ( x,t ) \Delta t ) \delta ( \xi )+\mathcal{O}(\Delta t^2).
\end{IEEEeqnarray*}
So the PDF of the whole propagator \(\Delta X_t\) is

\begin{IEEEeqnarray*}{rCl}
    p_{\Delta X}( \xi |\Delta t;x,t )& = & p_j( \xi |\Delta t;x,t ) *p_d( \xi |\Delta t;x,t )+\mathcal{O}(\Delta t^2)
    \\
    &=                        & \int_{\mathbb{R}}{p_j( \xi -\zeta |\Delta t;x,t ) p_d( \zeta |\Delta t;x,t ) d\zeta}+\mathcal{O}(\Delta t^2)
    \\
    &=                 & \int_{\mathbb{R}}{
                \left( w( \xi -\zeta |x,t ) \lambda ( x,t ) \Delta t+( 1-\lambda ( x,t ) \Delta t ) \delta ( \xi -\zeta ) \right) \frac{1}{\sqrt{2\pi b( x,t ) \Delta t}}\exp \left( -\frac{( \zeta -a( x,t ) \Delta t ) ^2}{2b( x,t ) \Delta t} \right)d\zeta}+\mathcal{O}(\Delta t^2)
    \\
    &=                        & \lambda ( x,t ) \Delta t\int_{\mathbb{R}}{w( \xi -\zeta |x,t ) \frac{\exp \left( -\frac{( \zeta -a( x,t ) \Delta t ) ^2}{2b( x,t ) \Delta t} \right)}{\sqrt{2\pi b( x,t ) \Delta t}}d\zeta}
     \\
     &&\<+\frac{1-\lambda ( x,t ) \Delta t}{\sqrt{2\pi b( x,t ) \Delta t}}\int_{\mathbb{R}}{\delta ( \xi -\zeta ) \exp \left( -\frac{( \zeta -a( x,t ) \Delta t ) ^2}{2b( x,t ) \Delta t} \right) d\zeta}+\mathcal{O}(\Delta t^2)
    \\
    &=                        & \lambda ( x,t ) \Delta t\int_{\mathbb{R}}{w( \xi -\zeta |x,t ) \frac{\exp \left( -\frac{( \zeta -a( x,t ) \Delta t ) ^2}{2b( x,t ) \Delta t} \right)}{\sqrt{2\pi b( x,t ) \Delta t}}d\zeta}
     +\frac{1-\lambda ( x,t ) \Delta t}{\sqrt{2\pi b( x,t ) \Delta t}}\exp \left( -\frac{( \xi -a( x,t ) \Delta t ) ^2}{2b( x,t ) \Delta t} \right)+\mathcal{O}(\Delta t^2)
    \\
    &=                        & \lambda ( x,t ) \Delta t\int_{\mathbb{R}}{\left[ w( \xi |x,t ) +\sum_{n=1}^{\infty}{\frac{( -\zeta ) ^n}{n!}\frac{\partial ^n}{\partial \xi ^n}w( \xi |x,t )} \right] \frac{\exp \left( -\frac{( \zeta -a( x,t ) \Delta t ) ^2}{2b( x,t ) \Delta t} \right)}{\sqrt{2\pi b( x,t ) \Delta t}}d\zeta}
    \\
    &&\< +\frac{1-\lambda ( x,t ) \Delta t}{\sqrt{2\pi b( x,t ) \Delta t}}\exp \left( -\frac{( \xi -a( x,t ) \Delta t ) ^2}{2b( x,t ) \Delta t} \right)+\mathcal{O}(\Delta t^2)
    \\
    &=                        & w( \xi |x,t ) \lambda ( x,t ) \Delta t+\frac{1-\lambda ( x,t ) \Delta t}{\sqrt{2\pi b( x,t ) \Delta t}}\exp \left( -\frac{( \xi -a( x,t ) \Delta t ) ^2}{2b( x,t ) \Delta t} \right)
    \\
    &&\< +\lambda ( x,t ) \Delta t\sum_{n=1}^{\infty}{\frac{( -1 ) ^n}{n!}\frac{\partial ^n}{\partial \xi ^n}w( \xi |x,t ) \int_{\mathbb{R}}{\zeta ^n\frac{\exp \left( -\frac{( \zeta -a( x,t ) \Delta t ) ^2}{2b( x,t ) \Delta t} \right)}{\sqrt{2\pi b( x,t ) \Delta t}}d\zeta}}+\mathcal{O}(\Delta t^2)
    \\
    &=                        & w( \xi |x,t ) \lambda ( x,t ) \Delta t+\frac{1-\lambda ( x,t ) \Delta t}{\sqrt{2\pi b( x,t ) \Delta t}}\exp \left( -\frac{( \xi -a( x,t ) \Delta t ) ^2}{2b( x,t ) \Delta t} \right)
    \\
    &&\< -\lambda ( x,t ) \sum_{k=1}^{\infty}{\frac{a^{2k-1}( x,t )}{( 2k-1 ) !}\frac{\partial ^n}{\partial \xi ^n}w( \xi |x,t ) a( x,t ) \Delta t^{2k}}
    \\
    &&\< +\lambda ( x,t ) \sum_{k=1}^{\infty}{\sum_{j=1}^k{\frac{( 2j-1 ) !!a^{2k-2j}( x,t ) b^{j}( x,t )}{( 2j ) !( 2k-2j ) !}}\frac{\partial ^n}{\partial \xi ^n}w( \xi |x,t ) \Delta t^{2k-j+1}}+\mathcal{O}(\Delta t^2)
    \\
    &=                        & w( \xi |x,t ) \lambda ( x,t ) \Delta t+\frac{1-\lambda ( x,t ) \Delta t}{\sqrt{2\pi b( x,t ) \Delta t}}\exp \left( -\frac{( \xi -a( x,t ) \Delta t ) ^2}{2b( x,t ) \Delta t} \right) +\mathcal{O}(\Delta t^2)
\end{IEEEeqnarray*}
\qed
\subsection{Proof of Lemma \ref{lem:kme}}\label{app:lem:kme}
\begin{IEEEeqnarray*}{Cl}
    p_X( x,t+\Delta t )                  & =\int_{\mathbb{R}}{p_X( x-\xi ,t ) p_{\Delta X}( \xi |\Delta t;x-\xi ,t ) d\xi}
    \\
    \frac{\partial p_X(x,t)}{\partial t} & =\lim_{\Delta t\rightarrow 0} \frac{p_X( x,t+\Delta t ) -p_X( x,t )}{\Delta t}
    \\
    & =\lim_{\Delta t\rightarrow 0} \frac{\displaystyle\int_{\mathbb{R}}{p_X( x-\xi ,t ) p_{\Delta X}( \xi |\Delta t;x-\xi ,t ) d\xi}-p_X( x,t )}{\Delta t}
    \\
    & \xlongequal{( \ref{KMexp} )}\lim_{\Delta t\rightarrow 0} \frac{\displaystyle\int_{\mathbb{R}}{\sum_{n=1}^{\infty}{\frac{( -\xi ) ^n}{n!}\frac{\partial ^n}{\partial x^n}\left[ p_X( x,t ) p_{\Delta X}( \xi |\Delta t;x,t ) \right]}d\xi}}{\Delta t}
    \\
    & =\lim_{\Delta t\rightarrow 0} \frac{1}{\Delta t}\sum_{n=1}^{\infty}{\frac{( -1 ) ^n}{n!}\frac{\partial ^n}{\partial x^n}\left[ p_X( x,t ) \int_{\mathbb{R}}{\xi ^np_{\Delta X}( \xi ;x,t,h ) d\xi} \right]}
    \\
    & =\sum_{n=1}^{\infty}{\frac{( -1 ) ^n}{n!}\frac{\partial ^n}{\partial x^n}\left[ p_X( x,t ) B_n( x,t ) \right]}
\end{IEEEeqnarray*}
where \((\ref{KMexp})\) comes from the series expansion
\begin{IEEEeqnarray}{c}\label{KMexp}
    p_X( x-\xi ,t ) p_{\Delta X}( \xi |\Delta t;x-\xi ,t ) = p_X( x,t ) p_{\Delta X}( \xi |\Delta t;x,t )
    +\sum_{n=1}^{\infty}{\frac{( -\xi ) ^n}{n!}\frac{\partial ^n}{\partial x^n}\left[ p_X( x,t ) p_{\Delta X}( \xi |\Delta t;x,t ) \right]}.
\end{IEEEeqnarray}
\qed
\section{Proofs of theorems}\label{proof_thm}
\subsection{Proof of Theorem \ref{thm:kmc}}\label{app:thm:kmc}
\begin{IEEEeqnarray*}{rCl}
    B_n( x,t ) & = & \lim_{\Delta t\rightarrow 0} \frac{1}{\Delta t}\int_{\mathbb{R}}{\xi ^np_{\Delta X}( \xi |\Delta t;x,t ) d\xi}
    \\
    &=             & \lim_{\Delta t\rightarrow 0} \frac{1}{\Delta t}\int_{\mathbb{R}}{\xi ^nw( \xi |x,t ) \lambda ( x,t ) \Delta td\xi}
    +\frac{1}{\Delta t}\int_{\mathbb{R}}{\frac{\xi ^n( 1-\lambda ( x,t ) \Delta t )}{\sqrt{2\pi b( x,t ) \Delta t}}\exp \left( -\frac{( \xi -a( x,t ) \Delta t ) ^2}{2b( x,t ) \Delta t} \right) d\xi}
    \\
    &=             & \lambda ( x,t ) \int_{\mathbb{R}}{\xi ^nw( \xi |x,t ) d\xi}
    +\lim_{\Delta t\rightarrow 0}\frac{ 1-\lambda ( x,t ) \Delta t }{\Delta t}\int_{\mathbb{R}}{\frac{\xi ^n}{\sqrt{2\pi b( x,t ) \Delta t}}\exp \left( -\frac{( \xi -a( x,t ) \Delta t ) ^2}{2b( x,t ) \Delta t} \right) d\xi}
    \\
    &=             & 
    \lambda ( x,t ) w_n( x,t ) +
    \begin{cases}\displaystyle
        \lim_{\Delta t\rightarrow 0} ( 1-\lambda ( x,t ) \Delta t ) a( x,t ) ,&n=1                             \\\displaystyle
        \lim_{\Delta t\rightarrow 0} ( 1-\lambda ( x,t ) \Delta t ) ( a^2( x,t ) \Delta t+b( x,t ) ) ,&n=2     \\\displaystyle
        \lim_{\Delta t\rightarrow 0} ( 1-\lambda ( x,t ) \Delta t ) a^{2k+1}( x,t ) \Delta t^{2k},&n=2k+1 \\\displaystyle
        \lim_{\Delta t\rightarrow 0}  
                ( 1-\lambda ( x,t ) \Delta t ) 
                \sum_{j=1}^{k+1}{\frac{( 2k+2 ) !( 2j-1 ) !!}{( 2j ) !( 2k-2j+2 ) !}a^{2k-2j+2}( x,t ) b^j( x,t ) \Delta t^{2k-j+1}} 
              ,&n=2k+2                                                   \\
    \end{cases}
    \\
    &=             & \begin{cases}
        \lambda ( x,t ) w_1( x,t ) +a( x,t ) ,&n=1 \\
        \lambda ( x,t ) w_2( x,t ) +b( x,t ) ,&n=2 \\
        \lambda ( x,t ) w_n( x,t ) ,&n\geqslant 3. \\
    \end{cases}
\end{IEEEeqnarray*}
\qed
\subsection{Proof of Theorem \ref{thm:fpme}}\label{app:thm:fpme}
\begin{IEEEeqnarray*}{rcl}
    p( x,t+\Delta t ) & - & p( x,t ) \approx \int_{\mathbb{R}}{p( x-\xi ,t ) w( \xi |x-\xi ,t ) \lambda ( x-\xi ,t ) \Delta td\xi}
    \\
    &&\< +\int_{\mathbb{R}}{\frac{( 1-\lambda ( x-\xi ,t ) \Delta t ) p( x-\xi ,t )}{\sqrt{2\pi b( x-\xi ,t ) \Delta t}}\exp \left( -\frac{( \xi -a( x-\xi ,t ) \Delta t ) ^2}{2b( x-\xi ,t ) \Delta t} \right) d\xi}-p( x,t )
    \\
    &=                   & \int_{\mathbb{R}}{p( x-\xi ,t ) \lambda ( x-\xi ,t ) w( \xi |x-\xi ,t ) d\xi}\Delta t-p( x,t ) \lambda ( x,t ) \Delta t
    \\
    &&\< +\int_{\mathbb{R}}{( 1-\lambda ( x-\xi ,t ) \Delta t ) \frac{p( x-\xi ,t ) -p( x,t )}{\sqrt{2\pi b( x-\xi ,t ) \Delta t}}\exp \left( -\frac{( \xi -a( x-\xi ,t ) \Delta t ) ^2}{2b( x-\xi ,t ) \Delta t} \right) d\xi}
    \\
    &=                   & \int_{\mathbb{R}}{p( x-\xi ,t ) \lambda ( x-\xi ,t ) w( \xi |x-\xi ,t ) d\xi}\Delta t-p( x,t ) \lambda ( x,t ) \Delta t
    \\
    &&\< +\int_{\mathbb{R}}{( 1-\lambda ( x-\xi ,t ) \Delta t ) \sum_{n=1}^{\infty}{\frac{( -\xi ) ^n}{n!}\frac{\partial ^n}{\partial x^n}\left[ \frac{p( x,t )}{\sqrt{2\pi b( x,t ) \Delta t}}\exp \left( -\frac{( \xi -a( x,t ) \Delta t ) ^2}{2b( x,t ) \Delta t} \right) \right]}d\xi}
    \\
    &=                   & \int_{\mathbb{R}}{p( x-\xi ,t ) \lambda ( x-\xi ,t ) w( \xi |x-\xi ,t ) d\xi}\Delta t-p( x,t ) \lambda ( x,t ) \Delta t
    \\
    &&\< +\sum_{n=1}^{\infty}{\frac{( -1 ) ^n}{n!}\frac{\partial ^n}{\partial x^n}\left[ p( x,t ) \int_{\mathbb{R}}{\xi ^n\frac{\exp \left( -\frac{( \xi -a( x,t ) \Delta t ) ^2}{2b( x,t ) \Delta t} \right)}{\sqrt{2\pi b( x,t ) \Delta t}}d\xi} \right]}
    \\
    &&\< -\sum_{n=1}^{\infty}{\frac{( -1 ) ^n}{n!}\frac{\partial ^n}{\partial x^n}\left[ p( x,t ) \int_{\mathbb{R}}{\lambda ( x-\xi ,t ) \xi ^n\frac{\exp \left( -\frac{( a( x,t ) \Delta t ) ^2}{2b( x,t ) \Delta t} \right)}{\sqrt{2\pi b( x,t ) \Delta t}}d\xi} \right]}\Delta t
    \\
    &=                   & \int_{\mathbb{R}}{p( x-\xi ,t ) \lambda ( x-\xi ,t ) w( \xi |x-\xi ,t ) d\xi}\Delta t-p( x,t ) \lambda ( x,t ) \Delta t
    \\
    &&\< -\sum_{k=1}^{\infty}{\frac{1}{( 2k-1 ) !}\frac{\partial ^n}{\partial x^n}\left[ p( x,t ) a^{2k-1}( x,t ) \right]}\Delta t^{2k-1}
    \\
    &&\< +\sum_{k=1}^{\infty}{\sum_{j=1}^k{\frac{( 2j-1 ) !!}{( 2j ) !( 2k-2j ) !}\frac{\partial ^n}{\partial x^n}\left[ p( x,t ) a^{2k-2j}( x,t ) b^j( x,t ) \right]}}\Delta t^{2k-j}
    \\
    &&\< +\lambda ( x,t ) \sum_{k=1}^{\infty}{\frac{1}{( 2k-1 ) !}\frac{\partial ^n}{\partial x^n}\left[ p( x,t ) a^{2k-1}( x,t ) \right]}\Delta t^{2k}
    \\
    &&\< -\lambda ( x,t ) \sum_{k=1}^{\infty}{\sum_{j=1}^k{\frac{( 2j-1 ) !!}{( 2j ) !( 2k-2j ) !}\frac{\partial ^n}{\partial x^n}\left[ p( x,t ) a^{2k-2j}( x,t ) b^j( x,t ) \right]}}\Delta t^{2k-j+1}
    \\
    &&\< +\sum_{k=1}^{\infty}{\sum_{n=1}^{2k}{\frac{1}{( 2k-n+1 ) !n!}\frac{\partial ^{2k-n+1}}{\partial x^{2k-n+1}}\lambda ( x,t ) \frac{\partial ^n}{\partial x^n}\left[ p( x,t ) a^{2k+1}( x,t ) \right]}\Delta t^{2k+2}}
    \\
    &&\< -\sum_{k=1}^{\infty}{\sum_{n=1}^{2k-1}{\sum_{j=1}^k{
            \frac{( 2k ) !( 2j-1 ) !!}{( 2k-n ) !n!( 2j ) !( 2k-2j ) !}\frac{\partial ^{2k-n}}{\partial x^{2k-n}}\lambda ( x,t ) \frac{\partial ^n}{\partial x^n}\left[ p( x,t ) a^{2k-2j}( x,t ) b^j( x,t ) \right]                            
        }}\Delta t^{2k-j+1}}
    \\
    &=                   & \int_{\mathbb{R}}{p( x-\xi ,t ) \lambda ( x-\xi ,t ) w( \xi |x-\xi ,t ) d\xi}\Delta t-p( x,t ) \lambda ( x,t ) \Delta t
    \\
    &&\< -\frac{\partial}{\partial x}( a( x,t ) p( x,t ) ) \Delta t+\frac{1}{2}\frac{\partial ^2}{\partial x^2}( b( x,t ) p( x,t ) ) \Delta t+\mathcal{O}(\Delta t^2)
\end{IEEEeqnarray*}
\begin{IEEEeqnarray*}{rCl}
    \frac{\partial p( x,t )}{\partial t}&= & \lim_{\Delta t\rightarrow 0} \frac{p( x,t+\Delta t ) -p( x,t )}{\Delta t}
    \\
    &=                                     & \int_{\mathbb{R} }{p( x-\xi ,t ) \lambda ( x-\xi ,t ) w( \xi |x-\xi ,t ) d\xi}-p( x,t ) \lambda ( x,t )
     -\frac{\partial}{\partial x}( a( x,t ) p( x,t ) ) +\frac{1}{2}\frac{\partial ^2}{\partial x^2}( b( x,t ) p( x,t ) )
\end{IEEEeqnarray*}
\qed

\subsection{Proof of Theorem \ref{thm:KM_diff_h}}\label{app:thm:KM_diff_h}
\begin{IEEEeqnarray*}{rCl}
    \frac{d}{dt}h( X_t ) &=&-\frac{d}{dt}\int{p\log pdx}=-\int{\frac{\partial p}{\partial t}\log pdx}-\frac{d}{dt}\int{pdx}
    \\
    &=&-\int{\left[ \sum_{n=1}^{\infty}{\frac{( -1 ) ^n}{n!}\frac{\partial ^n}{\partial x^n}( B_np )} \right] \log pdx}
    \\
    &=&-\sum_{n=1}^{\infty}{\frac{( -1 ) ^n}{n!}\int{\frac{\partial ^n}{\partial x^n}( B_np ) \log pdx}}
    \\
    &=&-\sum_{n=1}^{\infty}{\frac{1}{n!}\int{B_np\frac{\partial ^n}{\partial x^n}\log pdx}}
    \\
    &=&-\int{B_1p\frac{\partial}{\partial x}\log pdx}-\frac{1}{2}\int{B_2p\frac{\partial ^2}{\partial x^2}\log pdx}-\sum_{n=3}^{\infty}{\frac{1}{n!}\int{B_np\frac{\partial ^n}{\partial x^n}\log pdx}}
    \\
    &=&-\int{B_1\frac{\partial}{\partial x}pdx}-\frac{1}{2}\int{B_2p\left[ \frac{1}{p}\frac{\partial ^2p}{\partial x^2}-\left( \frac{\partial \log p}{\partial x} \right) ^2 \right] dx}
    -\sum_{n=3}^{\infty}{\frac{1}{n!}\int{B_np\frac{\partial ^n}{\partial x^n}\log pdx}}
    \\
    &=&\int{p\frac{\partial}{\partial x}B_1dx}-\frac{1}{2}\int{p\frac{\partial ^2}{\partial x^2}B_2dx}+\frac{1}{2}\int{B_2p\left( \frac{\partial \log p}{\partial x} \right) ^2dx}
    -\sum_{n=3}^{\infty}{\frac{1}{n!}\int{B_np\frac{\partial ^n}{\partial x^n}\log pdx}}
    \\
    &=&\mathbb{E} \left[ \frac{\partial}{\partial x}B_1( X_t,t ) -\frac{\partial ^2}{\partial x^2}B_2( X_t,t ) \right] +\frac{1}{2}J_{B_{2,t}}( X_t )
    -\sum_{n=3}^{\infty}{\frac{1}{n!}\mathbb{E} \left[ B_n( X_t,t ) \frac{\partial ^n}{\partial x^n}\log p_{X_t}( X_t,t ) \right]}
\end{IEEEeqnarray*}
\qed
\subsection{Proof of Theorem \ref{thm:KM_diff_i}}\label{app:thm:KM_diff_i}
\begin{IEEEeqnarray*}{rCl}
    \frac{d}{dt}I( X_0;X_t ) &=&\frac{d}{dt}h( X_t ) -\frac{d}{dt}h( X_t|X_0 )
    \\
    &=&\mathbb{E} \left[ \frac{\partial}{\partial x_t}B_1( X_t,t ) -\frac{\partial ^2}{\partial x^2}B_2( X_t,t ) \right] -\mathbb{E} \left[ \left. \frac{\partial}{\partial x_t}B_1( X_t,t ) -\frac{\partial ^2}{\partial x^2}B_2( X_t,t ) \right|X_0 \right]
    \\
    &&\<+\frac{1}{2}J_{B_{2,t}}( X_t ) -\frac{1}{2}J_{B_{2,t}}( X_t|X_0 )-\sum_{n=3}^{\infty}{\frac{1}{n!}\mathbb{E} \left[ B_n( X_t,t ) \frac{\partial ^n}{\partial x_{t}^{n}}\log p_{X_t}( X_t,t ) \right]}
    \\
    &&\<+\sum_{n=3}^{\infty}{\frac{1}{n!}\mathbb{E} \left[ B_n( X_t,t ) \frac{\partial ^n}{\partial x_{t}^{n}}\log p_{X_t|X_0}( X_t|X_0 ) \right]}
    \\
    &=&-\frac{1}{2}J_{B_{2,t}}( X_0;X_t ) -\sum_{n=3}^{\infty}{\frac{1}{n!}\mathbb{E} \left[ B_n( X_t,t ) \frac{\partial ^n}{\partial x_{t}^{n}}\log \frac{p_{X_t}( X_t )}{p_{X_t|X_0}( X_t|X_0 )} \right]}
    \\
    &=&-\frac{1}{2}J_{B_{2,t}}( X_0;X_t ) -\sum_{n=3}^{\infty}{\frac{1}{n!}\mathbb{E} \left[ B_n( X_t,t ) \frac{\partial ^n}{\partial x_{t}^{n}}\log \frac{p_{X_t}( X_t ) p_{X_0}( X_0 )}{p_{X_0,X_t}( X_0,X_t )} \right]}
    \\
    &=&-\frac{1}{2}J_{B_{2,t}}( X_0;X_t ) -\sum_{n=3}^{\infty}{\frac{1}{n!}\mathbb{E} \left[ B_n( X_t,t ) \frac{\partial ^n}{\partial x_{t}^{n}}\log \frac{p_{X_0}( X_0 )}{p_{X_0|X_t}( X_0|X_t )} \right]}
    \\
    &=&-\frac{1}{2}J_{B_{2,t}}( X_0;X_t ) +\sum_{n=3}^{\infty}{\frac{1}{n!}\mathbb{E} \left[ B_n( X_t,t ) \frac{\partial ^n}{\partial x_{t}^{n}}\log p_{X_0|X_t}( X_0|X_t ) \right]}
\end{IEEEeqnarray*}
\qed
\subsection{Proof of Theorem \ref{thm:diff_h}}\label{app:thm:diff_h}
\begin{IEEEeqnarray*}{rCl}
    \frac{d}{dt}h( X_t )& = & -\int_{\mathbb{R}}{\left[ \frac{\partial}{\partial t}p( x,t ) \right] \log p( x,t ) dx}-\int_{\mathbb{R}}{\frac{\partial}{\partial t}p( x,t ) dx}
    \\
    &=                      & -\int_{\mathbb{R}}{\left[ \frac{\partial}{\partial t}p( x,t ) \right] \log p( x,t ) dx}
    \\
    &=                      & -\int_{\mathbb{R}}{\left[ \begin{array}{c}
                \int_{\mathbb{R}}{p( x-\xi ,t ) \lambda ( x-\xi ,t ) w( \xi |x-\xi ,t ) d\xi}-p( x,t ) \lambda ( x,t )                  \\
                -\frac{\partial}{\partial x}( a( x,t ) p( x,t ) ) +\frac{1}{2}\frac{\partial ^2}{\partial x^2}( b( x,t ) p( x,t ) ) \\
            \end{array} \right] \log p( x,t ) dx}
    \\
    &=                      & -\int_{\mathbb{R}}{\int_{\mathbb{R}}{p( x,t ) \lambda ( x,t ) w( \xi |x,t ) \log p( x+\xi ,t ) d\xi}dx}
     +\int_{\mathbb{R}}{p( x,t ) \lambda ( x,t ) \log p( x,t ) dx}
    \\
    &&\< +\frac{1}{2}J_{b_t}( X_t ) +\mathbb{E} \left[ \frac{\partial}{\partial x}a( X_t,t ) -\frac{1}{2}\frac{\partial ^2}{\partial x^2}b( X_t,t ) \right]
    \\
    &=                      & \int_{\mathbb{R}}{\int_{\mathbb{R}}{p( x,t ) \lambda ( x,t ) w( \xi |x,t ) \log \frac{p( x,t )}{p( x+\xi ,t )}d\xi}dx}
     +\frac{1}{2}J_{b_t}( X_t ) +\mathbb{E} \left[ \frac{\partial}{\partial x}a( X_t,t ) -\frac{1}{2}\frac{\partial ^2}{\partial x^2}b( X_t,t ) \right]
    \\
    &=&\frac{1}{2}J_{b_t}( X_t ) +\mathbb{E} \left[ \lambda ( X_t,t ) \log \frac{p_{X_t}( X_t )}{p_{X_t}( X_t+\varXi )} \right] +\mathbb{E} \left[ \frac{\partial}{\partial x}a( X_t,t ) -\frac{1}{2}\frac{\partial ^2}{\partial x^2}b( X_t,t ) \right]
\end{IEEEeqnarray*}
where the proof of the diffusion part
\begin{IEEEeqnarray*}{c}
    \frac{1}{2}J_{b_t}( X_t ) +\mathbb{E} \left[ \frac{\partial}{\partial x}a( X_t,t ) -\frac{1}{2}\frac{\partial ^2}{\partial x^2}b( X_t,t ) \right]
\end{IEEEeqnarray*}
is the same as that in proof of Theorem \ref{thm:KM_diff_h} in Appendix \ref{app:thm:KM_diff_h}.
\qed
\subsection{Proof of Theorem \ref{thm:diff_i}}\label{app:thm:diff_i}
\begin{IEEEeqnarray*}{rCl}
    \frac{d}{dt}I( X_0;X_t ) &= & \frac{d}{dt}h( X_t ) -\frac{d}{dt}h( X_t|X_0 )
    \\
    &=                          & -\int_{\mathbb{R}}{\left[ \int_{\mathbb{R}}{\lambda ( x_t-\xi ,t ) w( \xi |x_t-\xi ,t ) p_{X_t}( x_t-\xi ) d\xi} \right] \log p_{X_t}( x_t ) dx_t}
    \\
    &&\< +\int_{\mathbb{R}}{\lambda ( x_t,t ) p_{X_t}( x_t ) \log p_{X_t}( x_t ) dx_t}
    \\
    &&\< +\int_{\mathbb{R}}{\int_{\mathbb{R}}{\left[ \int_{\mathbb{R}}{\lambda ( x_t-\xi ,t ) w( \xi |x_t-\xi ,t ) p_{X_t|X_0}( x_t-\xi |x_0 ) p_{X_0}( x_0 ) d\xi} \right] \log p_{X_t|X_0}( x_t|x_0 ) dx_tdx_0}}
    \\
    &&\< -\int_{\mathbb{R}}{\int_{\mathbb{R}}{\lambda ( x_t,t ) p_{X_t|X_0}( x_t|x_0 ) p_{X_0}( x_0 ) \log p_{X_t|X_0}( x_t|x_0 ) dx_t}dx_0}-\frac{1}{2}J_{b_t}( X_0;X_t )
    \\
    &=                          & \int_{\mathbb{R}}{\int_{\mathbb{R}}{\lambda ( x_t,t ) w( \xi |x_t,t ) p_{X_t}( x_t ) \log \frac{p_{X_t}( x_t )}{p_{X_t}( x_t+\xi )}d\xi}dx_t}
    \\
    &&\< -\int_{\mathbb{R}}{\int_{\mathbb{R}}{\int_{\mathbb{R}}{\lambda ( x_t,t ) w( \xi |x_t,t ) p_{X_t|X_0}( x_t|x_0 ) p_{X_0}( x_0 ) \log \frac{p_{X_t|X_0}( x_t|x_0 )}{p_{X_t|X_0}( x_t+\xi |x_0 )}d\xi}dx_t}dx_0}-\frac{1}{2}J_{b_t}( X_0;X_t )
    \\
    &=                          & \int_{\mathbb{R}}{\int_{\mathbb{R}}{\lambda ( x_t,t ) w( \xi |x_t,t ) p_{X_t}( x_t ) \log \frac{p_{X_t}( x_t )}{p_{X_t}( x_t+\xi )}d\xi}dx_t}
    \\
    &&\< -\int_{\mathbb{R}}{\int_{\mathbb{R}}{\int_{\mathbb{R}}{\lambda ( x_t,t ) w( \xi |x_t,t ) p_{X_0,X_t}( x_0,x_t ) \log \frac{p_{X_0,X_t}( x_0,x_t )}{p_{X_0,X_t}( x_0,x_t+\xi )}d\xi}dx_t}dx_0}-\frac{1}{2}J_{b_t}( X_0;X_t )
    \\
    &=                          & \int_{\mathbb{R}}{\int_{\mathbb{R}}{\lambda ( x_t,t ) w( \xi |x_t,t ) p_{X_t}( x_t ) \log \frac{p_{X_t}( x_t )}{p_{X_t}( x_t+\xi )}d\xi}dx_t}
    \\
    &&\< -\int_{\mathbb{R}}{\int_{\mathbb{R}}{\int_{\mathbb{R}}{\left[ \begin{array}{c}
                        \lambda ( x_t,t ) w( \xi |x_t,t ) p_{X_0|X_t}( x_0|x_t ) p_{X_t}( x_t )                                 \\
                        \times \log \frac{p_{X_0|X_t}( x_0|x_t ) p_{X_t}( x_t )}{p_{X_0|X_t}( x_0|x_t+\xi ) p_{X_t}( x_t+\xi )} \\
                    \end{array} \right] d\xi}dx_t}dx_0}-\frac{1}{2}J_{b_t}( X_0;X_t )
    \\
    &=                          & \int_{\mathbb{R}}{\int_{\mathbb{R}}{\lambda ( x_t,t ) w( \xi |x_t,t ) p_{X_t}( x_t ) \log \frac{p_{X_t}( x_t )}{p_{X_t}( x_t+\xi )}d\xi}dx_t}
    \\
    &&\< -\int_{\mathbb{R}}{\int_{\mathbb{R}}{\int_{\mathbb{R}}{\lambda ( x_t,t ) w( \xi |x_t,t ) p_{X_0|X_t}( x_0|x_t ) p_{X_t}( x_t ) \log \frac{p_{X_0|X_t}( x_0|x_t )}{p_{X_0|X_t}( x_0|x_t+\xi )}d\xi}dx_t}dx_0}
    \\
    &&\< -\int_{\mathbb{R}}{\int_{\mathbb{R}}{\int_{\mathbb{R}}{\lambda ( x_t,t ) w( \xi |x_t,t ) p_{X_0,X_t}( x_0,x_t ) \log \frac{p_{X_t}( x_t )}{p_{X_t}( x_t+\xi )}d\xi}dx_t}dx_0}-\frac{1}{2}J_{b_t}( X_0;X_t )
    \\
    &=                          & \int_{\mathbb{R}}{\int_{\mathbb{R}}{\int_{\mathbb{R}}{\lambda ( x_t,t ) w( \xi |x_t,t ) p_{X_0|X_t}( x_0|x_t ) p_{X_t}( x_t ) \log \frac{p_{X_0|X_t}( x_0|x_t )}{p_{X_0|X_t}( x_0|x_t+\xi )}d\xi}dx_t}dx_0}-\frac{1}{2}J_{b_t}( X_0;X_t )
    \\
    &=                          & \int_{\mathbb{R}}{p_{X_t}( x_t ) \int_{\mathbb{R}}{\lambda ( x_t,t ) w( \xi |x_t,t ) \int_{\mathbb{R}}{p_{X_0|X_t}( x_0|x_t ) \log \frac{p_{X_0|X_t}( x_0|x_t )}{p_{X_0|X_t}( x_0|x_t+\xi )}dx_0}d\xi}dx_t}-\frac{1}{2}J_{b_t}( X_0;X_t )
    \\
    &=                          & -\int_{\mathbb{R}}{p_{X_t}( x_t ) \int_{\mathbb{R}}{\lambda ( x_t,t ) w( \xi |x_t,t ) D( p_{X_0|X_t}\parallel p_{X_0|X_t,\varXi}|\varXi=\xi,X_t=x_t ) d\xi}dx_t}-\frac{1}{2}J_{b_t}( X_0;X_t )
    \\
    &=                          & -\mathbb{E} \left[ \lambda ( X_t,t ) D( p_{X_0|X_t}\parallel p_{X_0|X_{t,\varXi}}|X_t ) \right]  -\frac{1}{2}J_{b_t}( X_0;X_t ).
\end{IEEEeqnarray*}
\qed
\section{Proof of corollaries}\label{proof_coro}
\subsection{Proof of Corollary \ref{crl:add_PDF}}\label{app:crl:add_PDF}
The Kolmogorov equation of the state-homogeneous process is
\begin{IEEEeqnarray*}{c}
    \frac{\partial p( x,t )}{\partial t} =\lambda ( t ) \int_{\mathbb{R}}{p( x-\xi ,t ) w( \xi ,t ) d\xi}-\lambda ( t ) p( x,t ) -a( t ) \frac{\partial}{\partial x}p( x,t ) +\frac{b( t )}{2}\frac{\partial ^2}{\partial x^2}p( x,t ).
\end{IEEEeqnarray*}
Denote
\begin{IEEEeqnarray*}{rCl}
    P( k,t ) & =&\mathscr{F} \left[ p( x,t ) \right] =\int_{-\infty}^{+\infty}{p( x,t ) e^{-ikx}dx},
    \\W( k,t ) &=&\mathscr{F} \left[ w( x,t ) \right] =\int_{-\infty}^{+\infty}{w( x,t ) e^{-ikx}dx}
\end{IEEEeqnarray*}
to be the image function after Fourier transform.
Apply the Fourier transform to both sides, and we get
\begin{IEEEeqnarray*}{rCl}
    \frac{\partial}{\partial t}P( k,t ) & =&\lambda ( t ) P( k,t ) W( k,t ) -\lambda ( t ) P( k,t ) -ia( t ) kP( k,t ) -\frac{b( t ) k^2}{2}P( k,t )
    \\
    \frac{\partial}{\partial t}P( k,t ) & =&\left( \lambda ( t ) W( k,t ) -\lambda ( t ) -ia( t ) k-\frac{b( t ) k^2}{2} \right) P( k,t )
    \\
    P( k,t ) &=&P( k,0 ) \exp \left[ \int_0^t{\left( \lambda ( \tau ) W( k,\tau ) -\lambda ( \tau ) -ia( \tau ) k-\frac{b( \tau ) k^2}{2} \right) d\tau} \right].
\end{IEEEeqnarray*}

Apply the inverse Fourier transform and we get
\begin{align*}
        p(x,t)=\dfrac{1}{2\pi}\int_{\mathbb{R}}
        \left(
        \int_{\mathbb{R}} p(x,0)e^{-ikx} dx
        \right) \cdot \exp\left(
        \int_{0}^{t}\left[
        \left(
        \int_{\mathbb{R}} w(x,\tau)e^{-ikx}dx -1
        \right)\lambda(\tau)-ika(\tau)-\frac{k^2}{2}b(\tau)
        \right]d\tau +ikx
        \right)dk.
\end{align*}
\qed

\subsection{Proof of Corollary \ref{crl:diff_h_add}}\label{app:crl:diff_h_add}
\begin{IEEEeqnarray*}{rCl}
    \frac{d}{dt}h( X_t )& = &
    \frac{1}{2}J_{b_t}( X_t ) + \mathbb{E} \left[ \lambda ( X_t,t ) \log \frac{p_{X_t}( X_t )}{p_{X_t}( X_t+\varXi )} +\frac{\partial}{\partial x}a( X_t,t ) -\frac{1}{2}\frac{\partial ^2}{\partial x^2}b( X_t,t ) \right]
    \\
    & = &\frac{1}{2}J_{b_t}( X_t ) + \mathbb{E} \left[ \lambda (t ) \log \frac{p_{X_t}( X_t )}{p_{X_t}( X_t+\varXi )} +\frac{\partial}{\partial x}a( t ) -\frac{1}{2}\frac{\partial ^2}{\partial x^2}b( t ) \right]
    \\
    &=                      & \lambda ( t ) \mathbb{E} \left[ D( p_{X_t}\parallel p_{X_{t,\varXi}} ) \right] +\frac{1}{2}J_{b_t}( X_t )
\end{IEEEeqnarray*}
\qed
\end{document}